\documentclass[aps,prl,twocolumn,letterpaper,groupedaddress,10pt]{revtex4-2} 
\usepackage{amssymb,amsthm,amsmath,amsfonts}
\usepackage{graphicx,ulem,mathptmx}
\usepackage[shortlabels]{enumitem}
\usepackage[pdftex,dvipsnames,usenames]{xcolor}
\usepackage[colorlinks=true,urlcolor=blue,citecolor=blue,linkcolor=blue]{hyperref}

\begin{document}

\title{Statistical Benchmarking of Scalable Photonic Quantum Systems}

\author{J. Tiedau}
	\email{johannes.tiedau@uni-paderborn.de}
	\affiliation{Integrated Quantum Optics Group, Institute for Photonic Quantum Systems (PhoQS), Paderborn University, Warburger Stra\ss{}e 100, 33098 Paderborn, Germany}
\author{M. Engelkemeier}
	\affiliation{Integrated Quantum Optics Group, Institute for Photonic Quantum Systems (PhoQS), Paderborn University, Warburger Stra\ss{}e 100, 33098 Paderborn, Germany}
\author{B. Brecht}
	\affiliation{Integrated Quantum Optics Group, Institute for Photonic Quantum Systems (PhoQS), Paderborn University, Warburger Stra\ss{}e 100, 33098 Paderborn, Germany}
\author{J. Sperling}
	\email{jan.sperling@uni-paderborn.de}
	\affiliation{Integrated Quantum Optics Group, Institute for Photonic Quantum Systems (PhoQS), Paderborn University, Warburger Stra\ss{}e 100, 33098 Paderborn, Germany}
\author{C. Silberhorn}
	\affiliation{Integrated Quantum Optics Group, Institute for Photonic Quantum Systems (PhoQS), Paderborn University, Warburger Stra\ss{}e 100, 33098 Paderborn, Germany}

\begin{abstract}
	Targeting at the realization of scalable photonic quantum technologies, the generation of many photons, their propagation in large optical networks, and a subsequent detection and analysis of sophisticated quantum correlations are essential for the understanding of macroscopic quantum systems.
	In this experimental contribution, we explore the joint operation of all mentioned ingredients.
	We benchmark our time-multiplexing framework that includes a high-performance source of multiphoton states and a large multiplexing network, together with unique detectors with high photon-number resolution, readily available for distributing quantum light and measuring complex quantum correlations.
	Using an adaptive approach that employs flexible time bins, rather than static ones, we successfully verify high-order nonclassical correlations of many photons distributed over many modes.
	By exploiting the symmetry of our system and using powerful analysis tools, we can analyze correlations that would be inaccessible by classical means otherwise.
	In particular, we produce on the order of ten photons and distribute them over 64 modes.
	Nonclassicality is verified with correlation functions up to the 128th order and statistical significances of up to 20 standard deviations.
\end{abstract}

\date{\today}
\maketitle

\paragraph*{Introduction.---}

	We are in the midst of the second quantum revolution \cite{DM03}.
	That is, quantum systems are increasingly recognized as a means to establish next-generation information technologies \cite{KLM01,KMNRDM07,K08,RL09,BFV09,S09,SP19}.
	Currently, several physical platforms compete for advancing fundamental research to such practical implementations, including superconducting, atom- and ion-based, and photonic systems.
	Scalability, i.e., going from quantum processing with a few to many qubits, is one of the main concerns when it comes to assessing the future success of any realizations;
	see, e.g., Refs. \cite{HSSP99,VKLNFFMDS13,MMLLPB14,Betal17,ZPHKBKGGM17} for experiments in different large-scale quantum systems.

	Proof-of-concept studies often require the generation of complex forms of entanglement \cite{YULF08,CMP14,RMJFT14,GSVCRTF15,GSVCRTF16,Wetal18,Getal19,TTF19}.
	However, nonclassicality, as introduced in \cite{G63,G63ibid,M86,MW95}, can be an equally valuable resource for quantum information applications in photonic systems \cite{YBTNGK18,SLR19}.
	For example, sophisticated types of entanglement can be obtained when propagating single-mode nonclassical states in optical networks \cite{KSBK02,W02,VS14}.
	Furthermore, applications in quantum metrology can significantly benefit from nonclassical correlations as well \cite{KTVJ19}, such as achieved when propagating high-photon-number states in interferometers and detecting them with single-photon counters \cite{HB93,MZCSSDPB16}.
	Also, fundamental quantum interference phenomena can be studied via nonclassical photon states, e.g., through the generation of cat states \cite{OJTG07}.

	Because of the demands of quantum information science and the widespread availability of photonic systems in experiments, it is not surprising that studying quantum effects in optical scenarios has recently gained enormous momentum.
	In particular, quantum-optical interference \cite{KBMW98} of many photons in many modes appeals to the demand of scalability;
	see, e.g., Refs. \cite{Metal13,TTSSGHNSW15,RFMWK16,MJMTBKW17,AKJMSHRWJ17,NWXC18,NDBMTSGJS20} for recent theoretical and experimental investigations.
	For this purpose, and because of the importance for fundamental and applied sciences, a number of nonclassicality criteria for photon-number-based quantum correlations in multimode systems have been established and successfully applied in experiments over many decades \cite{M79,A88,AT92,SRV05,SV06,BAZPA07,ACLMWS08,ALCS10,MBWLN10,AOB12,SVA13,SBVHBAS15,APHM16,SCEMRKNLGVAW17,BQVC19,SPBTEWLNLGVASW20}.

	In addition, the scaling behavior of nonclassical correlations is vital for the future success of optical platforms.
	For example, boson sampling was among the first theoretical proposals that demonstrate quantum enhancements of quantum processors over classical systems \cite{AA11,AA13}, which was followed by experiments \cite{Setal13,BFRDARW13,TDHNSW13,CORBGSVMMS13,Zetal20}.
    This protocol exactly utilized the resources mentioned previously, many photons that are distributed over many modes and measured with photon counters \cite{Zetal18,Hetal19}.
	The core idea is that, for systems that are large enough, classical simulations are infeasible because of the exponentially growing demand for resources.
	Again, it was proven that quantum-optical nonclassicality is a prerequisite for the function of boson sampling \cite{RRC16,SLR17}.
	To answer how one can classically assess if a quantum protocol was successful, different benchmarks have been developed and applied to overcome the impractical increase of classical resources by identifying statistical signatures of quantum enhancement rather than looking at individual outcomes \cite{Betal14,Cetal14,AGKE15,WKUMTRB16,Getal18,PWRWTS19}.
	Still, the general question remains open if and how nonclassical correlations can be certified in experiments beyond boson sampling when the system size is subject to exponential scaling.

\begin{figure}
	\includegraphics[width=\columnwidth]{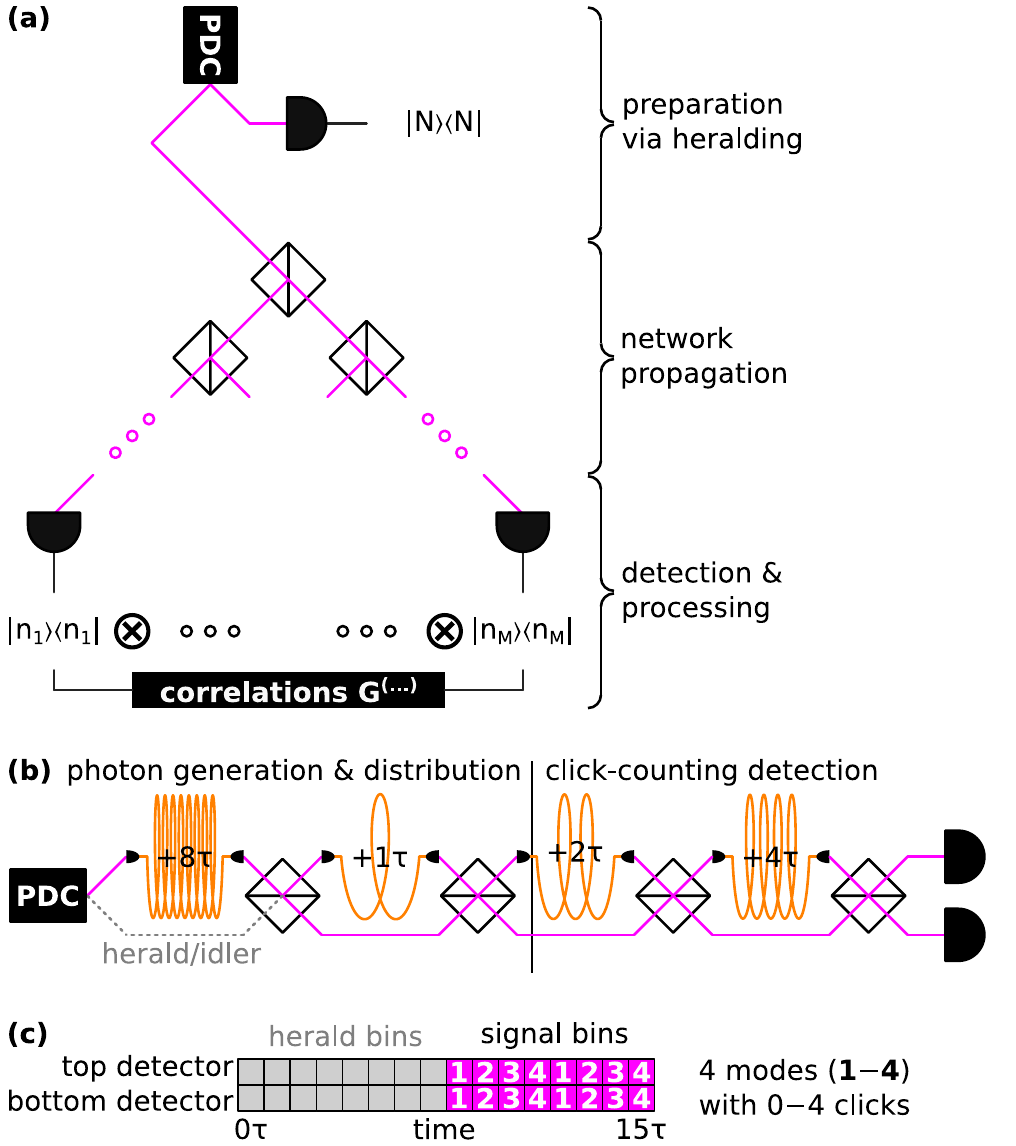}
	\caption{
		(a) Benchmark protocol in its spatial representation.
		Probing the future readiness of our system consists of three steps:
		the heralded generation of high-photon-number ($N$) states via a high-performance parametric down-conversion (PDC) source, the propagation of many photons in optical networks with many modes ($M$), and the photon-number-resolved detection to determine high-order correlations within and between the modes.
		(b) Efficient implementation as a time-bin multiplexing setup.
		Instead of spatial modes, we employ time bins that are separated by using ``$+n\tau$'' fiber-loop delay lines.
		Two on-off detectors, which are superconducting nanowire single-photon detectors in our case, record incident photons.
		At different levels, the multiplexing network can be separated (vertical line) to achieve a certain number of network modes, $M$, and a desired photon-number resolution via multiplexing of $D$ single-photon counters \cite{PTKJ96,BW03,SVA12true}.
		(c) Assigned time bins.
		With this separation approach, each time bin can be assigned to a network mode, and the number $D$ of possible joint clicks per mode identifies the order of correlation functions that are thereby accessible.
		Note that, in our experiment, the number of network and detection modes for the signal is much larger, $MD=128$, than in the depicted case, where $MD=16$.
	}\label{fig:Outline}
\end{figure}

    In this contribution, we devise and implement a benchmark protocol (Fig. \ref{fig:Outline}) to address the scaling problem for verifying quantum correlations in large photonic systems, including many photons, a highly multimode network for a quantum-coherent propagation, and single-photon counters.
    Rather than analyzing individual outputs with a computational demand that exceeds today's capabilities, our statistical approach renders it possible to provide insight into joint quantum correlations between all modes while being fine-grained enough to provide details of correlations between parts of the full system, which was not possible to date.
    Hence, high-order nonclassical correlations are verified with high significance using our directly accessible multimode nonclassicality criteria.
	In contrast to earlier studies, we benchmark all vital components of the experiment jointly---not separately---to characterize the scaling of nonclassicality.
	While it is not our intent to present a fully fleshed, universal quantum computer in this Letter, we can nonetheless certify the potential of our time-multiplexed optical quantum systems as a platform that operates reliably with highly correlated nonclassical light beyond the limitations of classical computational resources.

\paragraph{Benchmark protocol.---}

	For benchmarking the quantum characteristics of a full system, we devise a test with which we characterize nonclassicality via multimode correlations;
	see Fig. \ref{fig:Outline} for our implementation.
	For example, suppose a source produces $N$-photon states, $|N\rangle$ with $N\gg1$.
	A lossless and noise-free optical network distributes these photons over $M$ modes, where $M\gg1$.
	Then, one obtains the output
	\begin{equation}
		\label{eq:State}
		|\Psi_{N,M}\rangle
		=\sum_{\substack{n_1,\ldots,n_{M}\in\mathbb N: \\ n_1+\cdots+n_{M}=N}}
		\left[\frac{1}{M^N}\frac{N!}{n_1!\cdots n_{M}!}\right]^{1/2}
		|n_1,\ldots,n_{M}\rangle,
	\end{equation}
	being a highly entangled state.
	Afterward, we measure the joint photon-number distribution.
	For the state in Eq. \eqref{eq:State}, this results in correlation functions of the form
	\begin{equation}
	\begin{aligned}
		\label{eq:CorrFct}
		&G_{0}^{(m_1,\ldots,m_M)}
		=\langle{:}
			\hat n_1^{m_1}\cdots \hat n_M^{m_M}
		{:}\rangle
		\\
		=&
		\frac{1}{M^{m_1+\cdots+m_M}}
		\frac{N!}{(N-[m_1+\cdots+m_M])!},
	\end{aligned}
	\end{equation}
	for $m_1+\cdots+m_M\leq N$, and $G_{0}^{(m_1,\ldots,m_M)}=0$ otherwise, when assuming ideal detectors with full photon-number resolution, indicated by the subscript zero.
	From these correlations, we can infer the nonclassical correlations in one output mode or, more interestingly, over $K\leq M$ output modes.

	Our experiment does follow this idea but is clearly restricted by imperfections.
	Most importantly, we measure coincidences using $D$ detection bins and single-photon detectors, limiting our photon resolution.
	Rather than relying on the time-consuming reconstruction strategies, e.g., to obtain the perfect correlations in Eq. \eqref{eq:CorrFct}, a click-counting-based theory was developed \cite{SVA12true}.
	For the purpose of identifying $K$-mode nonclassical correlations, we use the matrix $\Gamma_K$ of moments that can be directly obtained from the measured click-counting coincidence statistics \cite{SVA13} by substituting $G_0^{(m_1,\ldots,m_M)}$ with $G^{(m_1,\ldots,m_M)}=\langle{:}\hat\pi_1^{m_1}\cdots\hat\pi_M^{m_M}{:}\rangle$, where $\hat\pi_j$ takes over the role of $\hat n_j$ \cite{SVA12true}, and using
	\begin{equation}
		\label{eq:MoM}
		\Gamma_K=\left(G^{(m_1+m'_1,\ldots,m_K+m'_K,0,\ldots,0)}\right)_{(m_1,\ldots,m_K),(m'_1,\ldots,m'_K)}.
	\end{equation}
	The rows and columns of $\Gamma_K$ are identified through the multi-indices $(m_1,\ldots,m_K)$ and $(m'_1,\ldots,m'_K)$, respectively, that can take values $m_j,m_j'\in\{0,\ldots,D/2\}$ when measuring up to $D$th-order correlations in the $j$th mode.
	For the $K$-mode correlations under consideration, we trace over the remaining $M-K$ modes, i.e., taking zeroth moments ($\hat \pi^0=\hat1$) only.
	For $K$-mode classically correlated light, we have a positive semidefinite matrix of moments, $\Gamma_K\geq0$.
	Conversely, whenever the minimal eigenvalue of $\Gamma_K$ is less than zero, nonclassical $K$-mode correlations are certified \cite{SBVHBAS15}.

	The results of the described analysis then demonstrate the quantum performance of the full system:
    the higher the nonclassicality, the better the quantum characteristics of the source, the propagation in the network, and the subsequent measurement for verifying quantum features.
	The distribution of the nonclassical states results in the expectation that quantum correlations should increase with increasing $K$ since an ever-increasing fraction of all available multimode correlations are accessed when $K$ approaches $M$.
    Exceeding the specific states in Eq. \eqref{eq:State}, the click-counting-based approach can be applied to general inputs and extends to networks beyond the homogeneous splitting configuration that we use here, significantly broadening the application of the proposed method.

\paragraph{Scaling behavior and exploiting symmetry.---}

	In addition to the experimental challenge of having a compatible source of quantum light and a nonclassicality-preserving spreading in the optical network from which the ultimate output photons have to be detected sufficiently well, a hindrance that is overcome by our benchmark protocol is an exponential scaling in the data processing.
	Even when focusing on the signal alone, i.e., ignoring the herald for the time being, we would need to record $2^{128}\sim10^{38}$ different counting patterns for each possible scheme of coincidence clicks from the $MD=128$ time bins for the two employed single-photon detectors ($\sim 10^{14}$ Ybyte of data).
	Clearly, this is infeasible on commonly accessible, classical computers.

	We exploit the expected symmetry of this system under exchanges of modes [cf. Eqs. \eqref{eq:State} and \eqref{eq:CorrFct}] to reduce this number significantly, leaving us with an exponentially decreased value of $129$ distinguishable counting patterns, $0$--$128$ clicks from the signal bins.
	This enables us to restrict ourselves to the number of joint clicks for our statistical benchmarking.
	To account for deviations from the symmetry, we measure the single-count rates from the individual time bins and assign a systematic uncertainty by analyzing the uniformity of the single count rates over the detection bins.
	This systematic error estimated from our datasets is in the range $4\%$--$7\%$.
	See the Supplemental Material for details \cite{SuppMat}.
	It is worth emphasizing that symmetry and nonclassicality are unrelated concepts;
	classical states---such as a mixture of $M$-mode coherent states $\int dP(\alpha)(|\alpha\rangle\langle\alpha|)^{\otimes M}$ for a probability distribution $P$---can be symmetric as well.
	Thus, this approach does not introduce any bias in the data processing.
	Moreover, when applied to inhomogeneous networks, this symmetry-based approach can be applied by averaging over all outputs at the expense of losing this particular information.
	Note that such a mixing cannot introduce nonclassicality either.

	Another essential point of the scaling behavior is that data processing becomes impractical, again, because of the exponential increase of information with the number of quantum systems.
	For determining the minimal eigenvalue of the correlation matrix $\Gamma_K$ in Eq. \eqref{eq:MoM}, we have to analyze this matrix which consists of
	\begin{equation}
		\left(\frac{D}{2}+1\right)^K
	\end{equation}
	columns and rows.
	This results from the previously mentioned fact that each row (likewise, column) for each of the $K$ modes consists of correlation functions with orders ranging from $0$ to $D/2$ \cite{SVA13}.
	For example, a $\sim 10^{19}\times10^{19}$ matrix of moments $\Gamma_K$ has to be processed for $D=2$ and $K=M=64$.
	Again, the mode-exchange symmetry can be used to overcome this exponential scaling for our benchmark (see Supplemental Material \cite{SuppMat} for technical details), leaving us with a maximal size of $65\times65$ for the probing the overall nonclassicality of our system.

\begin{figure*}
	\includegraphics[width=\textwidth]{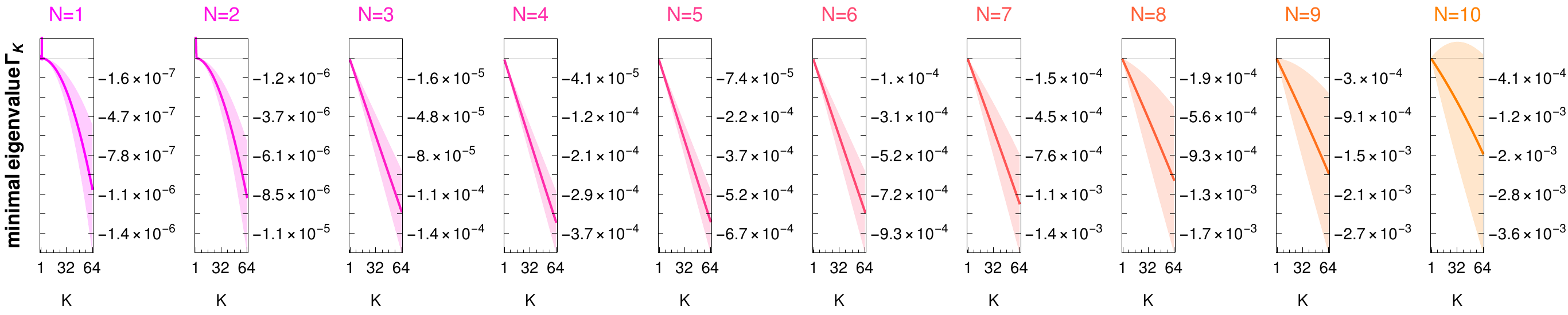}
	\caption{
		Benchmark of $K$-mode nonclassicality via negative minimal eigenvalue of the matrix of moments $\Gamma_K$ for $N\in\{1,\ldots,10\}$ photons distributed over $M=64$ modes.
		The classical-quantum boundary, zero, is shown as a gray horizontal line.
		The pump power for the PDC process is $500\,\mathrm{\mu W}$, and the adaptive time-bin width is twice the width of the Gaussian pulse width (i.e., $2\sigma$) after propagation through fibers.
		One standard-deviation error margin is shown as shaded area and includes random and systematic uncertainties.
		Mean values and random errors are obtained from raw coincidence counts, without any correcting postprocessing for imperfections.
		For most of the 640 analyzed cases shown here, nonclassicality is verified and increases with the number of photons $N$ and modes $K$ between which correlations are probed.
	}\label{fig:resultplot}
\end{figure*}

\paragraph{Experiment and adaptive time bins.---}

    To explore the challenging regime of large quantum states spread over many modes, we use type-II parametric down-conversion in a periodically poled potassium titanyl phosphate waveguide.
    This source is pumped with pulsed light from a Ti:sapphire laser at a repetition rate of $50\,\mathrm{kHz}$ and has shown single-mode emission \cite{ECMS11, Getal11} at a wavelength of around $1\,550\,\mathrm{nm}$, as well as high brightness values \cite{HBLNGS16}, being critical for this experiment.
    Detecting bright quantum states is an even more challenging problem.
    To achieve photon-number resolution, various approaches in the few photon regime have been shown that involve, for example, transition-edge sensors and time multiplexing \cite{LMN08,ASSBW03,FJPF03}.
    We expand these approaches and show time-multiplexing beyond $100$ discrete bins, with a time separation $\tau=100\,\mathrm{ns}$, while maintaining a high, uniform bin efficiency.
    Our custom-designed time-multiplexing unit features high average transmission values of $86.1\%$ and $81.3\%$ (first and second input modes, respectively) in combination with a high bin uniformity ($\lesssim 7\%$ variation over all bins) and no intrinsic cross talk between time bins.
    These unique specifications render our demanding study possible.
    Finally, both output ports from the time multiplexing are connected to superconducting nanowire single-photon detectors, with a dead time $\sim60\,\mathrm{ns}$, which can record joint click patterns over the time bins.

    For analyzing the resulting coincidence counts, one typically employs static time bins for extracting coincidence counts.
    Our following analysis, however, shows that it is beneficial to consider a dynamic binning instead.
    This is due to the dispersion in optical fibers, which result in a broadening of the pulses of light that carry the photons.
    A static time bin becomes too narrow because a prolonged propagation in a dispersive fiber causes significant broadening, resulting in too few photons in the relatively narrow coincidence time-bin window.
    Conversely, a too broad static coincidence window leads to many counts that stem from uncorrelated background noise, decreasing the impact of signal photons in earlier time bins.
    This issue can be resolved by considering an adaptive scheme in which the binning is chosen according to the width of the traveling pulses of light, being related to approaches from Refs. \cite{PCVZB12,MFL13,NRBU20}.
    With this advanced binning strategy, the relative width of coincidence windows for all time bins remains constant with respect to the pulse width, rather than having an unfavorable constant absolute width.
    Fitting the signals in individual time bins with Gaussian distributions, we found pulse widths ranging from $70\,\mathrm{ps}$ to $700\,\mathrm{ps}$, depending on the amount of dispersive broadening and lower bounded by the combined detector and readout jitter.

\paragraph*{Results and discussion.---}

	In Fig. \ref{fig:resultplot}, the results for one measured dataset are shown.
	We discuss up to $N=10$ photons that have been distributed over $M=64$ modes.
	The depicted results show the nonclassicality in terms of a negative minimal eigenvalue of the matrix $\Gamma_K$ for benchmarking quantum correlations between $1\leq K\leq M$ modes.

	The values depicted as solid curves in Fig. \ref{fig:resultplot} are in agreement with our intuition that the nonclassicality increases with increasing photon number;
	please mind the different scaling of the vertical axes.
	Also, the more modes are correlated, i.e., increasing $K$, the higher the verified nonclassical correlations, which is maximized when considering all modes, $K=M=64$.
	Because of impurities in the generation of our heralded photons states, the nonclassicality also varies with $N$.
	For instance, for $N=1$ and $N=2$, the single-mode case $K=1$ does not exhibit nonclassicality as the accumulated noise contributions are too high when tracing over all other $M-1$ modes.
	For large photon numbers, e.g., $N=10$, the error bars are too high to make statistically meaningful statements;
	this effect further increases for even higher photon numbers, and is therefore not depicted here.
	For the chosen pump power, the states with $N$ from three to seven exhibit the highest quality of nonclassicality when including error margins.

	This sample of our vast data analysis (see Supplemental Material \cite{SuppMat} for additional studies) demonstrates a successful benchmark concerning our system's scalability.
	Namely, without performing any correcting postprocessing of our data and rigorously including random and systematic uncertainties, we are able to achieve the goal of our proposed protocol to certify the nonclassical features of our system together with their functional dependencies that resemble ideal theoretical predictions.

\begin{figure}[b]
	\includegraphics[width=.8\columnwidth]{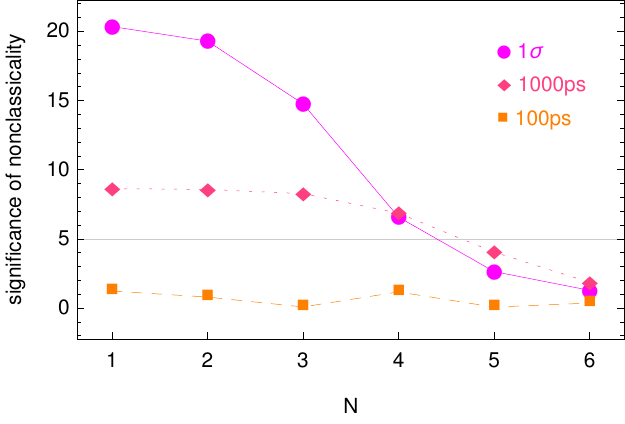}
	\caption{
		Significance of verified nonclassicality of heralded $N$-photon states for a pump power of $150\,\mathrm{\mu W}$.
		An adaptive binning with $1\sigma$ pulse width is compared to static time bins with $100\,\mathrm{ps}$ and $1\,000\,\mathrm{ps}$.
		A 5 standard-deviation significance threshold is shown as horizontal line for reference.
		For this source characterization, we have $128$ detection bins, allowing us to determine single-mode (i.e., $M=1$) nonclassicality with moments up to the $128$th order.
	}\label{fig:signplot}
\end{figure}

    While we focused on the spreading of quantum correlations in a network so far, we also characterize the source and detection system alone, i.e., the case $M=1$.
	In Fig. \ref{fig:signplot}, we particularly study the impact of the dynamically chosen coincidence windows when compared to static ones.
	Therein, the shown statistical significance is the distance of the minimal eigenvalue of the single-mode matrix of moments, including correlations up to the $128$th order, to the classical bound zero in units of the error margin.
	We observe that the adaptive binning is, in general, favorable for detecting nonclassicality when compared to commonly applied static bins.
	A $100\,\mathrm{ps}$ window---approximating the minimum pulse width $\sigma_{\min}=70\,\mathrm{ps}$---is too narrow for later, thus dispersed, pulses, negatively affecting the number of recorded coincidences.
	Conversely, a $1\,000\,\mathrm{ps}$ window---relating to $\sigma_{\max}=700\,\mathrm{ps}$---is too broad for earlier pulses, thus recording too much of uncorrelated background light.
	In both cases, we find a diminished significance of the probed $N$-photon states compared to the adaptive binning.
	The latter adaptive approach renders it possible to certify nonclassicality with a significance up to $\sim20$ standard deviations for the dataset under study.

	It is worth mentioning that, in comparison with Fig. \ref{fig:resultplot}, the experiment in Fig. \ref{fig:signplot} is carried out for a smaller pump power of the PDC process, resulting in a lower $N$.
	A comprehensive analysis of all binnings and pump powers, $15\,\mathrm{\mu W}$--$1000\,\mathrm{\mu W}$, can be found in the Supplemental Material \cite{SuppMat}, including a detailed discussion of data processing, error estimation, and a joint nonclassicality characterization between signal and idler \cite{SBVHBAS15}.
	In addition, theoretical studies about heralding of multiphoton states, including the impact of higher photon-number contributions and losses of the heralding detectors, can be found in Refs. \cite{SVA14,ELDBDPSS20}.
	Again, we emphasize that our data have not been corrected for those or any other imperfections, and our nonclassicality analysis is purely based on the sampled moments from the measured raw click-counting statistics.

\paragraph*{Conclusion.---}

	In summary, we experimentally analyzed the scalability of a photonic quantum system by benchmarking the joint operation of source, optical network, and single-photon counters.
	For assessing the nonclassical features of our system, we probed correlations between the output modes when distributing heralded multiphoton states obtained from a PDC source.
	In this way, we analyzed up to 64 quantum correlated modes for up to ten photon states with correlation functions with a maximal order of 128.
	By exploiting the symmetry of our system, we were able to reduce the problem of analyzing a $10^{19}\times10^{19}$ matrix of moments---being infeasible on a classical computer---to a problem of a $65\times 65$ matrix with the same information content.
	Thus, and despite the exponential scaling, nonclassicality has been verified with a statistical significance of up to 20 standard deviations.
	This was achieved by an adaptive approach to defining time-bin modes, superseding conventional static binnings.
	Furthermore, our results of our quantum correlation analysis have been directly obtained from the measured data, without performing any postprocessing to correct for unwanted impurities.

    Our approach is based on a detection theory for signals that are uniformly distributed multiple detection bins and measured single-photon detectors.
    Recent works show that the balanced splitting can be overcome \cite{BQVC19} and that arbitrary detectors can be employed in this scheme as well \cite{SCEMRKNLGVAW17}, allowing for a detector-independent generalization of our technique.
    Moreover, we implemented a resource-friendly time-multiplexed version of our protocol with a uniform spreading of photon-number states in a network.
    However, such a restriction does not apply for the underlying data processing strategy, which addresses the exponential scaling and certifies multimode nonclassical correlations, as it can accept other input states and other network configurations, realized in various optical settings.
    This capacity to advance our benchmark protocol together with the broad applicability demonstrates the future potential of our approach for fundamental and applied investigations in quantum sciences and technology.

	The Integrated Quantum Optics group acknowledges financial support through the European Commission through the ERC project QuPoPCoRN (Grant No. 725366) and the H2020-FETFLAG-2018-03 project PhoG (Grant No. 820365) and funding through the Gottfried Wilhelm Leibniz-Preis (Grant No. SI1115/3-1).


\end{document}


\title{Supplemental Material}

\author{J. Tiedau}
	\affiliation{Integrated Quantum Optics Group, Institute for Photonic Quantum Systems (PhoQS), Paderborn University, Warburger Stra\ss{}e 100, 33098 Paderborn, Germany}
\author{M. Engelkemeier}
	\affiliation{Integrated Quantum Optics Group, Institute for Photonic Quantum Systems (PhoQS), Paderborn University, Warburger Stra\ss{}e 100, 33098 Paderborn, Germany}
\author{B. Brecht}
	\affiliation{Integrated Quantum Optics Group, Institute for Photonic Quantum Systems (PhoQS), Paderborn University, Warburger Stra\ss{}e 100, 33098 Paderborn, Germany}
\author{J. Sperling}
	\affiliation{Integrated Quantum Optics Group, Institute for Photonic Quantum Systems (PhoQS), Paderborn University, Warburger Stra\ss{}e 100, 33098 Paderborn, Germany}
\author{C. Silberhorn}
	\affiliation{Integrated Quantum Optics Group, Institute for Photonic Quantum Systems (PhoQS), Paderborn University, Warburger Stra\ss{}e 100, 33098 Paderborn, Germany}

\begin{abstract}
	Here, we provide additional technical details to complement the key findings reported in the main text.
	In Sec. \ref{supp.sec:IdealCase}, a calculation for ideal scenario is provided.
	In Sec. \ref{supp.sec:Theory}, the theoretical description for higher-order, multimode correlation functions, including nonclassicality criteria, are reviewed.
	In Sec. \ref{supp.sec:Implementation}, a more detailed description of the experimental setup is given.
	In Sec. \ref{supp.sec:SystematicError}, systematic errors are discussed.
	In Sec. \ref{supp.sec:JointCorrelations}, joint nonclassical correlations between signal and idler are analyzed for all data sets.
	In Sec. \ref{supp.sec:StaticVsDynamic}, the impact of selecting static and dynamic time bins is studied.
	In Sec. \ref{supp.sec:HeraldedSignificances}, an extended nonclassicality analysis of heralded states is provided.
\end{abstract}

\date{\today}
\maketitle
\appendix

\section{Ideal photodetection scenario}\label{supp.sec:IdealCase}

    In the lossless and noise-free case, the input state $|N\rangle=\hat a^{\dag N}|\mathrm{vac}\rangle/\sqrt{N!}$ is spread over $M$ modes.
    Assuming a uniform distribution, the bosonic creation operator $\hat a^\dag$ can be expressed in terms of the output modes as $\hat a^\dag=(\hat a_1^\dag+\cdots+\hat a_M^\dag)/\sqrt{M}$.
    This results in the output state as given in the main text,
    \begin{equation}
    \begin{aligned}
        &|\Psi_{N,M}\rangle=\frac{\left(\hat a_1^\dag+\cdots+\hat a_M^\dag\right)^N}{\sqrt{M^N N!}}|\mathrm{vac}\rangle
        \\
        =&\frac{1}{\sqrt{M^NN!}}\sum_{\substack{n_1,\ldots,n_{M}\in\mathbb N: \\ n_1+\cdots+n_{M}=N}}
		\frac{N!}{n_1!\cdots n_{M}!}\hat a_1^{\dag n_1}\cdots \hat a_M^{\dag n_M}
		|\mathrm{vac}\rangle
        \\
        =&\sum_{\substack{n_1,\ldots,n_{M}\in\mathbb N: \\ n_1+\cdots+n_{M}=N}}
		\sqrt{\frac{1}{M^N}\frac{N!}{n_1!\cdots n_{M}!}}
		|n_1,\ldots,n_{M}\rangle.
    \end{aligned}
    \end{equation}

    For computing ideal photon-number correlations, one can, for example, consider the generating function
    \begin{equation}
    \begin{aligned}
        &\tilde g(z_1,\ldots,z_M)=\langle\Psi_{N,M}|z_1^{\hat n_1}\cdots z_M^{\hat n_M}|\Psi_{N,M}\rangle
        \\
        =&\sum_{\substack{n_1,\ldots,n_{M}\in\mathbb N: \\ n_1+\cdots+n_{M}=N}}
		\frac{1}{M^N}\frac{N!}{n_1!\cdots n_{M}!}
		z_1^{n_1}\cdots z_M^{n_M}
		\\
		=&\frac{\left(z_1+\cdots+z_M\right)^N}{M^N},
    \end{aligned}
    \end{equation}
    where $\hat n_j=\hat a_j^\dag\hat a_j$ is the photon-number operator for the $j$th mode.
    Applying normal ordering, ${:}e^{(z-1)\hat n}{:}=z^{\hat n}$, the normally ordered powers of the photon-number operator can be conveniently expressed as $\partial_z^m{:}e^{(z-1)\hat n}{:}|_{z=1}={:}\hat n^m{:}$.
    Thus, ideal multimode moments in the main text are straightforwardly obtained via derivatives of the generating function,
    \begin{equation}
    \begin{aligned}
        &\left.\partial_{z_1}^{m_1}\cdots\partial_{z_M}^{m_M}\tilde g(z_1,\ldots,z_M)\right|_{z_1=\cdots=z_M=1}
        \\
        =&\frac{N!}{(N-[m_1+\cdots+m_M])!}\frac{\left(1+\cdots+1\right)^{N-[m_1+\cdots+m_N]}}{M^N}
        \\
        =&\frac{N!}{(N-[m_1+\cdots+m_M])!}\frac{1}{M^{m_1+\cdots+m_N}}
    \end{aligned}
    \end{equation}
    for $m_1+\cdots+m_M\leq N$ and zero otherwise.

\section{Click-counting detection theory and higher-order, multimode nonclassicality criteria}\label{supp.sec:Theory}

	For our statistical analysis, we utilize a previously devised method for theoretically describing the finite photon-number-resolution and the resulting correlation functions \cite{SVA12true,SVA13}.
	With this method, the click-counting statistics for an ensemble of $D$ on-off detectors in a multiplexing configuration for each mode is given by the positive operator-valued measure defined through the elements
	\begin{equation}
		\hat\Pi_n={:}
			\binom{D}{n}
			\hat\pi^{n}\left[\hat 1-\hat\pi\right]^{D-n}
		{:},
	\end{equation}
	where $0\leq n\leq D$ is the number of clicks, 
	and the expectation value of ${:}\hat\pi{:}$ yields the probability that a single detector produces a click, including imperfections.
	Note that the click-counting description approaches the standard photoelectric detection model in the limit of many detectors $D$ and for low photon numbers \cite{SVA12true}.

	From the joint detection of multiple modes, $M$-mode correlation functions can be inferred \cite{SVA13}.
	They can be obtained from a recorded click distribution as
	\begin{equation}
	\begin{split}
		&G^{(m_1,\ldots,m_M)}
		=\langle{:}\hat\pi_1^{m_1}\cdots\hat\pi_M^{m_M}{:}\rangle
		\\
		=&\sum_{n_1,\ldots,n_M=0}^D\frac{\binom{n_1}{m_1}\cdots\binom{n_M}{m_M}}{\binom{D}{m_1}\cdots\binom{D}{m_M}}
		\underbrace{\langle\hat\Pi_{n_1}\otimes\cdots\otimes\hat\Pi_{n_M}\rangle}_{=c(n_1,\ldots,n_M)},
	\end{split}
	\end{equation}
	where $\binom{x}{m}=0$ for $m>x$ and $0\leq m_j\leq D$ for each mode, $j=1,\ldots,M$.
	In this formula, $c(n_1,\ldots,n_M)$ is the joint click-counting distribution for $n_j$ clicks for the $j$th mode.
	Note that the moments simplify to $\langle{:}\hat\pi_1^{m_1}\cdots\hat\pi_M^{m_M}{:}\rangle=\langle{:}\hat\pi^{m_1+\cdots+m_M}{:}\rangle$ for a symmetric system under study.

	From these multimode correlation functions, nonclassicality criteria can be formulated in terms of matrices of moments \cite{SVA13}.
	Specifically, the matrix $\Gamma$ of (higher-order) multimode moments is given by
	\begin{equation}
		\Gamma=\left(G^{(m_1+m'_1,\ldots,m_M+m'_M)}\right)_{(m_1,\ldots,m_M),(m'_1,\ldots,m'_M)},
	\end{equation}
	where rows and columns are defined by the multi-indices $(m_1,\ldots,m_M)$ and $(m'_1,\ldots,m'_M)$, respectively, each taking values from $(0,\ldots,0)$ to $(D/2,\ldots,D/2)$.
	Then, the product $MD$ defines the maximal order of correlations which are used to obtain $\Gamma$; in our experiment, we have $MD=128$.
	For classical states, the matrix of moments satisfies
	\begin{equation}
		\label{eq:ClConstraint}
		\vec f^\dag \Gamma \vec f\stackrel{\text{cl.}}{\geq}0 \text{ for all vectors } \vec f.
	\end{equation}
	A violation of this constraint certifies nonclassicality \cite{SVA13}.
	In particular, we can choose $\vec f$ to be the normalized eigenvector to the minimal eigenvalue $\lambda$ of $\Gamma$, which has to be nonnegative eigenvalue for classical light \cite{SBVHBAS15}.

	Furthermore, and as introduced in the main text, we consider certain $K$-mode submatrices of $\Gamma$, indicated as $\Gamma_K$, to quantify the correlations between $1\leq K\leq M$ modes of the full multimode systems.
	Because of symmetry, where $G^{(m_1+m'_1,\ldots,m_K+m'_K,0,\ldots,0)}=G^{([m_1+\cdots+m_K]+[m'_1+\cdots+m'_K])}$ holds true, entries of the submatrix $\Gamma_K$ can occur several times.
	Let $d_m$ be the number of rows and columns for which $m_1+\cdots+m_K=m$, where $0\leq m\leq \kappa=KD/2$ for the rows and columns.
	Then, the submatrix of moments can be put in a block form of identical entries, 
	\begin{equation}
		\Gamma_K=\begin{pmatrix}
			G^{(0+0)}\vec n_{d_0}\vec n_{d_0}^\dag & \hdots & G^{(0+\kappa)}\vec n_{d_0}\vec n_{d_\kappa}^\dag
			\\
			\vdots & \ddots & \vdots
			\\
			G^{(\kappa+0)}\vec n_{d_\kappa}\vec n_{d_0}^\dag & \hdots & G^{(\kappa+\kappa)}\vec n_{d_\kappa}\vec n_{d_\kappa}^\dag
		\end{pmatrix},
	\end{equation}
	where $\vec n_d=[1,\ldots,1]^\mathrm{T}$ is a $d$-dimensional vector of ones.

	To find the minimal eigenvalue $\lambda$ to $\Gamma_K$ for the aforementioned nonclassicality condition, we can also put the corresponding eigenvector in a block form, $\vec f=[\vec f_0^\mathrm{\,T},\ldots,\vec f_\kappa^\mathrm{\,T}]^\mathrm{T}$.
	This yields the eigenvalue equation
	\begin{equation}
		\Gamma_K\vec f=\lambda\vec f
		\quad\Leftrightarrow\quad
		\begin{pmatrix}
			\gamma_0\vec n_{d_0} \\ \vdots \\ \gamma_\kappa\vec n_{d_\kappa}
		\end{pmatrix}
		=\lambda
		\begin{pmatrix}
			\vec f_{0} \\ \vdots \\ \vec f_{\kappa}
		\end{pmatrix},
	\end{equation}
	where $\gamma_j=\sum_{l=0}^\kappa G^{(j+l)}\vec n_{d_l}^\dag\vec f_l$.
	The eigenvalue equation then implies that $\vec f_j\propto \vec n_{d_j}$ holds true for nonzero eigenvalues.
	Specifically, we can set $\vec f_j=\tilde f_{j}\vec n_{d_j}/\sqrt{d_j}$ such that $\vec f_j^\dag\vec f_j=|\tilde f_j|^2$.
	We can then rewrite $\gamma_j=\sum_{l=0}^\kappa G^{(j+l)}\sqrt{d_l}\tilde f_l$ and define the smaller matrix $\tilde \Gamma_K=(G^{(j+l)})_{j,l\in\{0,\ldots,\kappa\}}$, the vector $\vec{\tilde f}=(\tilde f_j)_{j\in\{0,\ldots,\kappa\}}$, and a scaling matrix $S=\mathrm{diag}[\sqrt{d_0},\ldots,\sqrt{d_\kappa}]$.
	With these definitions, the initial eigenvalue problem can be rewritten as
	\begin{equation}
		\Gamma_K\vec f=\lambda\vec f
		\,\,\Leftrightarrow\,\,
		\tilde \Gamma_K S \vec{\tilde f}=\lambda S^{-1}\vec{\tilde f}
		\,\,\Leftrightarrow\,\,
		(S\,\tilde \Gamma_K S) \vec{\tilde f}=\lambda\vec{\tilde f},
	\end{equation}
	resulting in the minimal eigenvalue $\lambda$, i.e., the same outcome for the classicality constraint in Eq. \eqref{eq:ClConstraint}, $\vec f^\dag \Gamma_K \vec f=\sum_{j,l=0}^\kappa G^{(j+l)}\vec f_j^\dag\vec n_{d_j}\vec n_{d_l}^\dag\vec f_l=\sum_{j,l=0}^\kappa G^{(j+l)}[n_{d_j}n_{d_l}]^{1/2}\tilde f_j^\ast\tilde f_l =\vec{\tilde f}^\dag S\, \tilde\Gamma_K S\vec{\tilde f}$.
	Thus, we can study the rescaled and smaller matrix of moments $S\, \tilde\Gamma_K S$, without changing the outcome of our nonclassicality analysis.

	What is left is determining the multiplicities $d_m$ for $0\leq m\leq \kappa=KD/2$.
	Meaning that we want to determine how many combinations of $(m_1,\ldots,m_K)$ yield the same sum $m=m_1+\cdots+m_K$.
	For this purpose, we can apply a generating function approach,
	\begin{equation}
	\begin{split}
		g(z)
		=&\sum_{m_1=0}^{D/2}\cdots\sum_{m_K=0}^{D/2}z^{m_1+\cdots+m_K}
		=\left[\frac{1-z^{D/2+1}}{1-z}\right]^{K}
		\\
		=&\sum_{m=0}^\kappa d_m z^m
		\quad\Rightarrow\quad
		d_m=\frac{1}{m!}\left.\frac{\partial^m g(z)}{\partial z^m}\right|_{z=0},
	\end{split}
	\end{equation}
	showing that the sought-after parameters are the coefficients for the monomials $z^m$.
	Also note that the higher-order derivative can be rater efficiently determined through an iteration of first-order derivatives, $\frac{1}{m}\frac{\partial}{\partial z}\left[\frac{1}{(m-1)!}\frac{\partial^{m-1}}{\partial z^{m-1}}\right]=\frac{1}{m!}\frac{\partial^m}{\partial z^m}$.

\begin{figure*}
	\includegraphics[width=.95\textwidth]{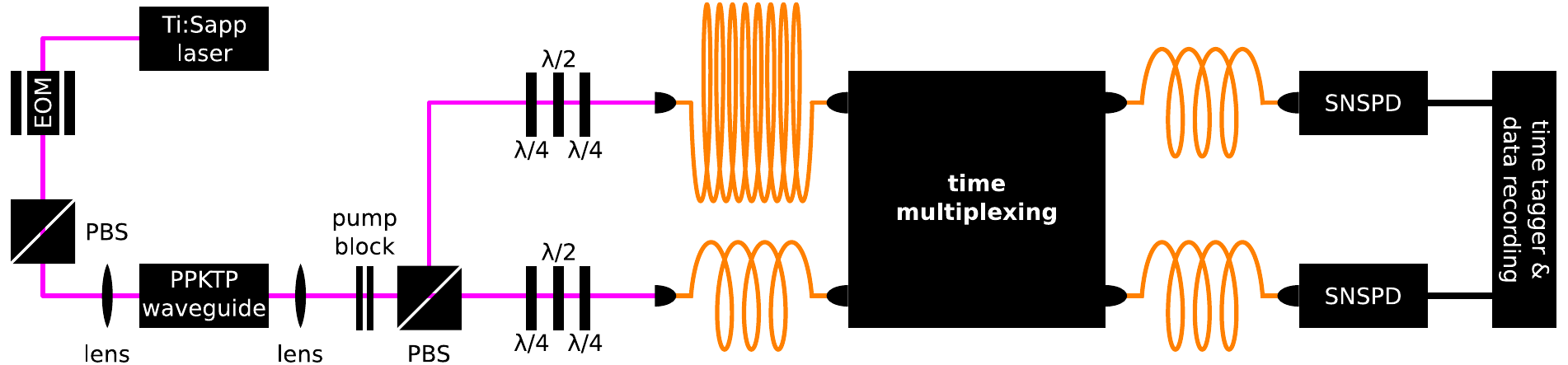}
	\caption{
		A pulse train from a Ti:Sapp laser is picked with an electro-optic modulator (EOM) and used as a pump for parametric-down conversion inside a periodically poled potassium titanyl phosphate (PPKTP) crystal.
		Signal and idler modes from the down-conversion process are separated by a polarizing beam splitter (PBS).
		Half-wave plates ($\lambda$/2) and quarter-wave plates ($\lambda$/4) are used to adjust the polarization for the time-multiplexing device, which is finally connected to the superconducting nanowire single-photon detectors (SNSPDs).
	}\label{supp.fig:Setup}
\end{figure*}

\section{Additional details on the experimental implementation}\label{supp.sec:Implementation}

	A schematic overview of our setup is shown in Fig. \ref{supp.fig:Setup}.
	A two-mode squeezed vacuum state is generated in a periodically poled potassium titanyl phosphate (PPKTP) crystal through a type-II parametric down-conversion process (PDC) \cite{ECMS11, Getal11}.
	The process is pumped with a Ti:Sapphire laser at $775\,\mathrm{nm}$ and a repetition rate of $76\,\mathrm{MHz}$.
	Using an electro-optic modulator (EOM) and a polarizing beam splitter (PBS), the repetition rate is down-sampled to a final measurement rate of $50\,\mathrm{kHz}$.
	In contrast to typical configurations, we do not apply any spectral filtering to the pump spectrum and the output state in order to increase the source brightness and efficiency.
	This leads to a spectrally multimode state with an expected spontaneous PDC bandwidth of $17\,\mathrm{nm}$, which does not derogate the results as only photon-number correlations are investigated.
	A silicon filter was inserted to block the $775\,\mathrm{nm}$ pump light after the PDC source.

	The two orthogonal polarizations, namely signal and idler (likewise, modes $A$ and $B$), from the PDC state are spatially separated with a PBS and adjusted with waveplates to couple each to the main axes of a polarization-maintaining fiber.
	Both fibers are connected to a custom-built time-multiplexing device (TMD) that separates each input mode into $128$ modes ($64$ time bins and two detection modes).
	Both output ports of the TMD are connected to superconducting nanowire single-photon detectors (SNSPDs).
	The time bins have a $100\,\mathrm{ns}$ time separation to account for the SNSPD dead time.
	In front of the SNSPDs, fiber polarization controllers maximize the detector efficiency with a value of around $90\%\pm 5\%$.
	The SNSPD response is measured with a time tagger and stored on a computer for further processing.
    Our TMD splits the signal into $128$ outputs in total.
    Thus, the number of network modes $M$ and detection bins $D$ is fixed, $MD=128$.
    To apply the click-counting data processing technique, we need at least two detection bins, $D=2$, which leaves us with maximally $M=64$ available network modes.

\begin{table}[b]
	\caption{
		Width of time-bin windows for a static binning (left) and an adaptive binning (right).
		The latter is given in units of the detected Gaussian pulse width $\sigma$ after propagation through the fibers.
		Note that, on average over all bins, the coincidence windows are comparable within each row of the table.
		The shortest and longest fiber paths led to pulse widths $\sigma_{\min}=70\,\mathrm{ps}$ and $\sigma_{\max}=700\,\mathrm{ps}$, respectively.
	}\label{supp.tab:Windows}
	\begin{tabular}{rr}
		\hline\hline
		\hspace{3cm}static & \hspace{3cm}dynamic
		\\
		\hline
		$20\,\mathrm{ps}$ & $0.1\sigma$
		\\
		$50\,\mathrm{ps}$ & $0.2\sigma$
		\\
		$100\,\mathrm{ps}$ & $0.5\sigma$
		\\
		$150\,\mathrm{ps}$ & $0.8\sigma$
		\\
		$200\,\mathrm{ps}$ & $1.0\sigma$
		\\
		$500\,\mathrm{ps}$ & $1.5\sigma$
		\\
		$1\,000\,\mathrm{ps}$ & $2.0\sigma$
		\\
		$1\,500\,\mathrm{ps}$ & $2.5\sigma$
		\\
		$2\,000\,\mathrm{ps}$ & $3.0\sigma$
		\\
		$2\,500\,\mathrm{ps}$ & $3.5\sigma$
		\\
		$5\,000\,\mathrm{ps}$ & $4.0\sigma$
		\\
		$6\,000\,\mathrm{ps}$ & $6.0\sigma$
		\\
		$8\,000\,\mathrm{ps}$ & $8.0\sigma$
		\\
		$10\,000\,\mathrm{ps}$ & $15.0\sigma$
		\\
		$15\,000\,\mathrm{ps}$ & $20.0\sigma$
		\\
		$30\,000\,\mathrm{ps}$ & $40.0\sigma$
		\\
		\hline\hline
	\end{tabular}

\end{table}

	Different time windows (see Table \ref{supp.tab:Windows}) are applied in the evaluation to optimize quantum efficiency and background count rate, which is nonnegligible because of the pulse-picking extinction ratio.
	We differentiate between static windows with a fixed time duration for all bins and dynamic windows that are adjusted for each individual bin.
	To adapt the windows for the latter, we fitted all 256 bins for signal and idler with Gaussian functions, where a multiple of the extracted standard deviation $\sigma$ is used define the dynamic window size.
	The shortest fiber path ($20\,\mathrm{m}$) had a standard deviation of $70\,\mathrm{ps}$, which is primarily limited by the overall detection jitter.
	In contrast, the longest fiber path ($2.5\,\mathrm{km}$) had a measured pulse distribution of $700\,\mathrm{ps}$, mainly defined by fiber dispersion.
	In addition, the measurement was carried out for various pump powers,
	\begin{equation}
		P\in\{15\,\mathrm{\mu W}, 50\,\mathrm{\mu W}, 150\,\mathrm{\mu W}, 500\,\mathrm{\mu W}, 1\,000\,\mathrm{\mu W}\}.
	\end{equation}
	Conditioning to a given signal or idler event realizes a heralding of a multiphoton quantum state in the remaining mode.

\section{Symmetry considerations for a systematic error estimation}\label{supp.sec:SystematicError}

	An assumption for our ideal model is that the intensity is distributed uniformly over all detection bins.
	In reality, however, there are slight deviations from this ideal case.
	Those impurities are used to estimate a systematic error.

\begin{figure}[b]
	\includegraphics[width=.8\columnwidth]{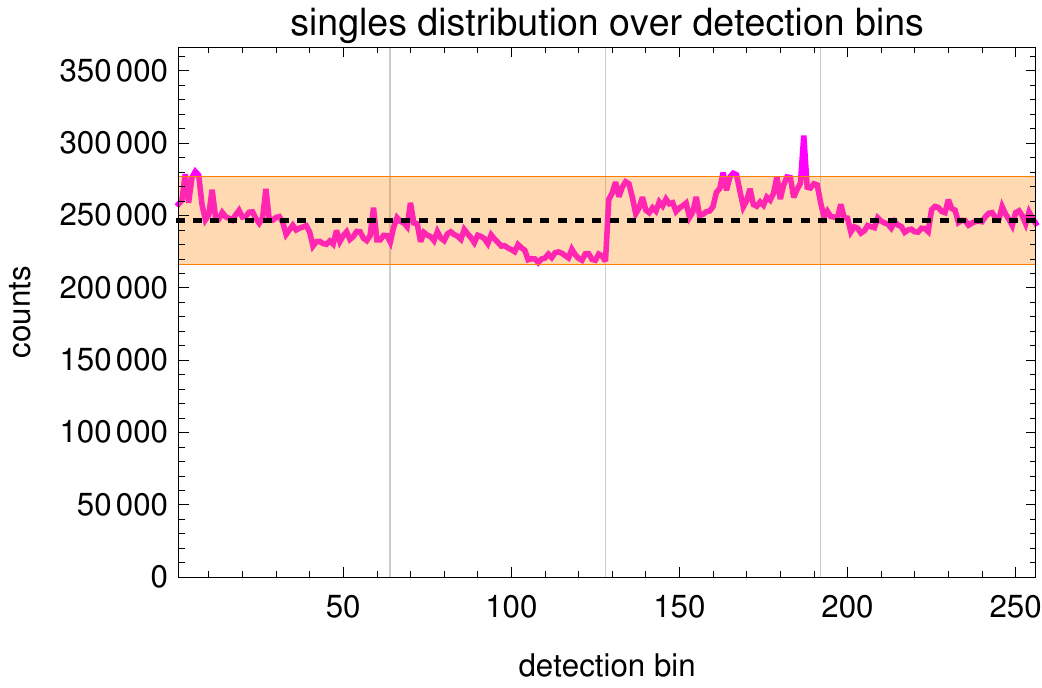}
	\caption{
		Single counts (magenta) from each detection bin for a pump power $P=500\,\mathrm{\mu W}$ and dynamic coincidence window of $3$ times the pulse width $\sigma$.
		The first, second, third, and fourth sectors (bin numbers $1$--$64$, $65$--$128$, $129$--$192$, and $193$--$256$, respectively) include the events from the first detector for the signal, the first detector for the idler, the second detector for the signal, and the second detector for the idler, respectively.
		The dashed horizontal line depicts the mean value and the orange region shows a $95\%$ confidence interval (i.e., two standard deviations), which relates to the systematic error to account for deviations from the symmetry assumption.
	}\label{supp.fig:DetectionBinSingles}
\end{figure}

	In Fig. \ref{supp.fig:DetectionBinSingles}, we exemplify one scenario of measured single counts from each detector for one given value of pump power $P$ and one choice of dynamic detection window.
	Therein, the first 64 detection bins are the time bins from first detector (i.e., the top detector in Fig. \ref{supp.fig:Setup}) for the signal light;
	the next 64 detection bins are those from the same detector but for the idler;
	and the last two collections of 64 detection bins each represent a similar separation for the second detector (bottom detector in Fig. \ref{supp.fig:Setup}) into signal and idler bins.

	The standard deviation in the counts determines the unwanted fluctuations over the detection bins.
	Scaling this variance to the mean number of single counts then provides an estimate for how much the premise of a uniform distribution is violated \cite{SCEMRKNLGVAW17,SPBTEWLNLGVASW20}, which is ${\sim}6\%$ for the scenario depicted in Fig. \ref{supp.fig:DetectionBinSingles}.
	This quantity also defines the relative systematic error to account for symmetry imperfection, which is obtained for each pump power and each detection window separately and combined with other statistical (i.e., random) uncertainties, resulting in the full error bars, as used throughout this work.

\section{Data sets and analysis of joint correlations}\label{supp.sec:JointCorrelations}

\begin{figure}[b]
	\includegraphics[width=.8\columnwidth]{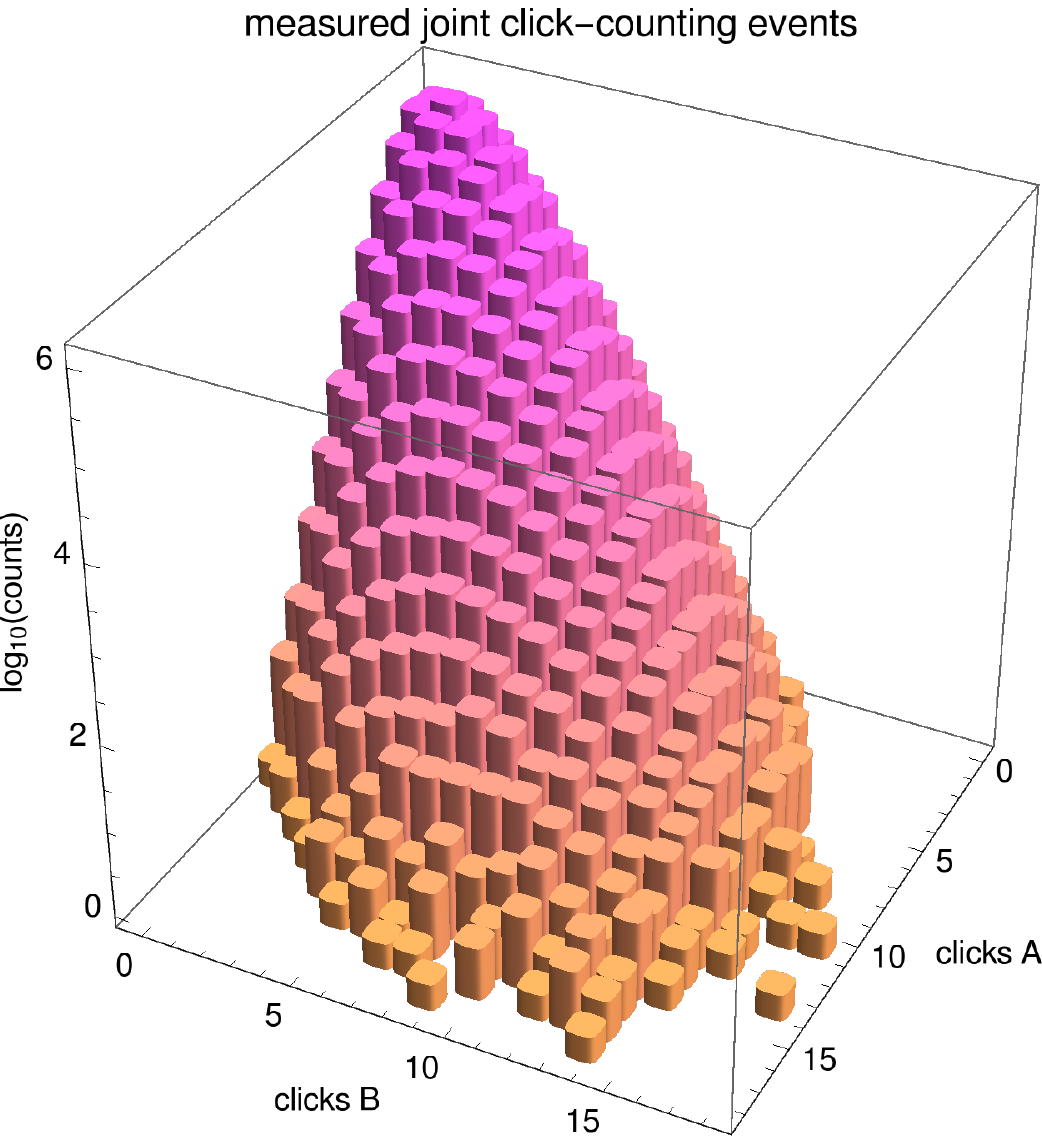}
	\caption{
		Example of recorded events for the pump power $P=500\,\mathrm{\mu W}$ and dynamic $3\sigma$ coincidence window, shown on a logarithmic scale as a function of the click numbers for signal and idler ($A$ and $B$).
		For click numbers larger than the depicted ones, and less than the maximum of 128, no events have been recorded.
	}\label{supp.fig:RawCounts}
\end{figure}

	In the following, we analyze the quality of our setup by studying the joint correlations between signal and idler \cite{SBVHBAS15}.
	A typical data set of the joint click-counting events is shown in Fig. \ref{supp.fig:RawCounts}.
	The events may be collected in the matrix $C(n_A,n_B)$, where $n_j$ denotes the number of clicks for the signal ($j=A$) and idler ($j=B$).
	Then, the joint click-counting probability can be estimated via $c(n_A,n_B)=C(n_A,n_B)/C$, where $C=\sum_{n_A,n_B} C(n_A,n_B)$ is the total number of events.
	As described in Sec. \ref{supp.sec:Theory}, this joint click-counting distribution can be used to construct a two-mode matrix of moments $\Gamma$ and determine its negativity via its minimal eigenvalue \cite{SBVHBAS15}.
	The highest order of joint moments is $128$ for $A$ and $B$, resulting in a $65^2\times65^2$ matrix $\Gamma$.

\begin{figure}[t]
	\includegraphics[width=.9\columnwidth]{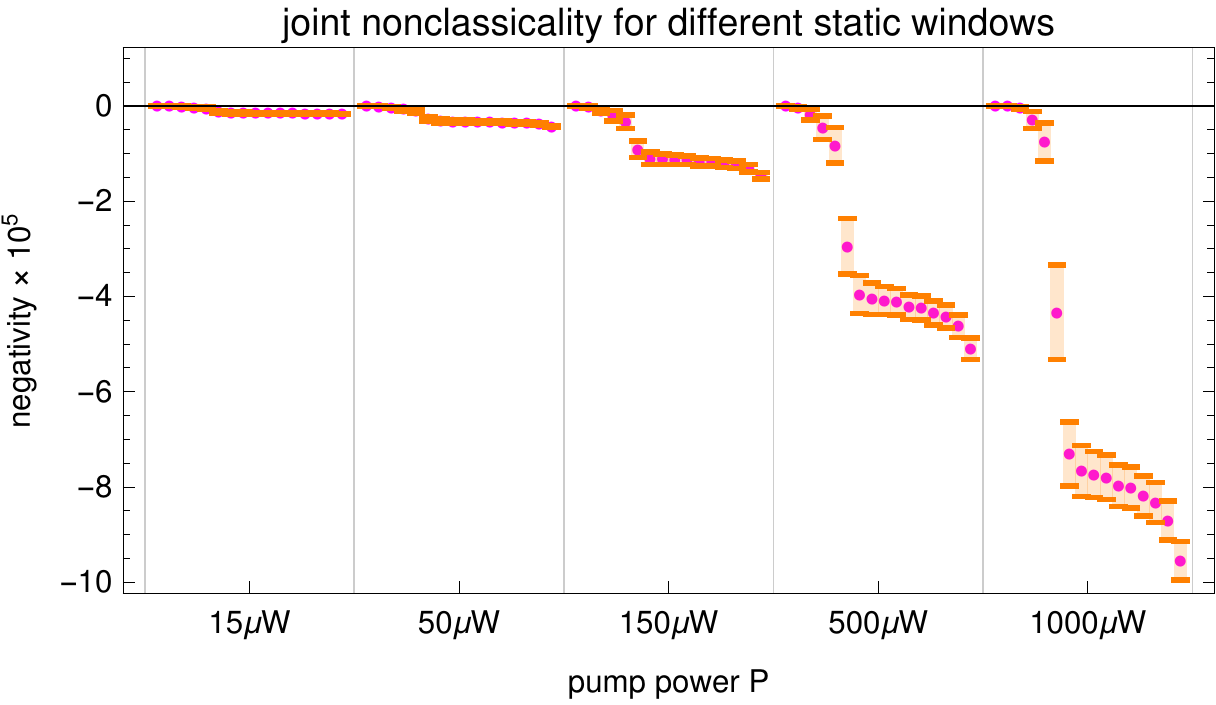}
	\\\vspace{2ex}
	\includegraphics[width=.9\columnwidth]{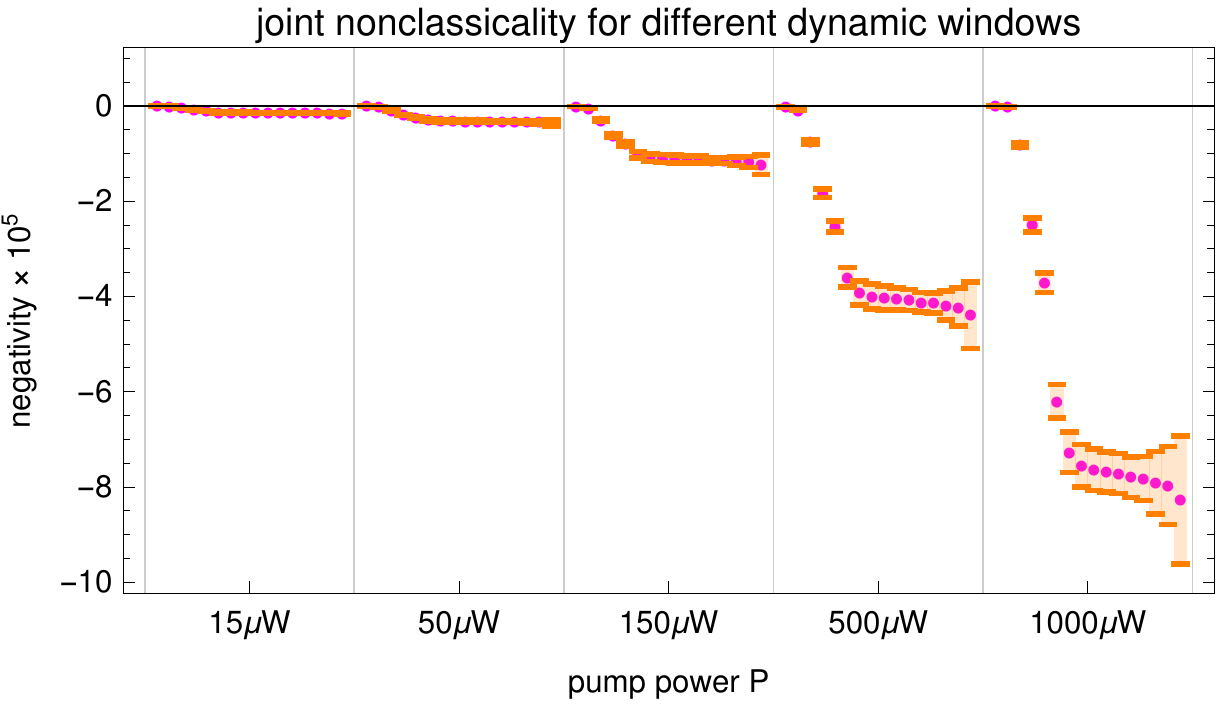}
	\caption{
		Highest detectable nonclassicality (magenta bullet points) from joint click-counting statistic as a function of pump power for static (top) and dynamic (bottom) coincidence windows, including error bars (orange) consisting of random and systematic uncertainties.
		Within each power setting (separated by vertical lines), the negativities given from left to right correspond to the coincidence windows listed in Table \ref{supp.tab:Windows} from top to bottom.
	}\label{supp.fig:JointNcl}
\end{figure}

	Figure \ref{supp.fig:JointNcl} depicts the resulting negativity of the matrix of moments $\Gamma$ for static and dynamic coincidence windows (top and bottom plot, respectively) for various pump powers.
	The plots include a $\pm1$-standard-deviation error margin which combines random errors of the determination of moments from the measured click-counting statistics, as well as systematic errors discussed in Sec. \ref{supp.sec:SystematicError}.
	One can clearly identify the nonclassical character of the joint signal-idler correlations which is certified with high statistical significance.

	Furthermore, and anticipating the discussion in the following Sec. \ref{supp.sec:StaticVsDynamic}, it can be seen that the chosen detection windows cover similar ranges of nonclassical correlations for the static and dynamic cases.
	However, the static scenario comes with significantly increased error bars.
	Compare, for example, static and dynamic windows for $P=1\,000\,\mathrm{\mu W}$, top and bottom plot in Fig. \ref{supp.fig:JointNcl}.
	There are points in which both exhibit comparable negativities (${\approx}-4\times 10^{-5}$), yet with rather different error margins attached to them, resulting in a verification of nonclassicality with a significance of ${\sim}4$ and ${\sim}18$ standard deviations for the static and dynamic case, respectively.
	This corresponds a more than four-fold increased performance when comparing static with dynamic time bins.

\section{Statistical analysis of static and dynamic binning}\label{supp.sec:StaticVsDynamic}

	As seen from the previous analysis and as discussed on a conceptional basis in the main text, one can observe major benefits from considering a dynamic binning when compared to a static one.
	The main idea is that a too narrow window decreases the number of coincidences whereas a too broad window increases the contribution of uncorrelated background noise.
	But the propagation in fibers of different lengths can significantly change the width $\sigma$ of the traveling pulses of light, meaning that a static window either collects too much background for shorter travel times (narrower pulses) or does not collect enough events for longer propagation times  because of the ever increasing pulse width.
	Here, let us confirm this intuitive picture with measured data.

\begin{figure}[t]
	\includegraphics[width=.475\columnwidth]{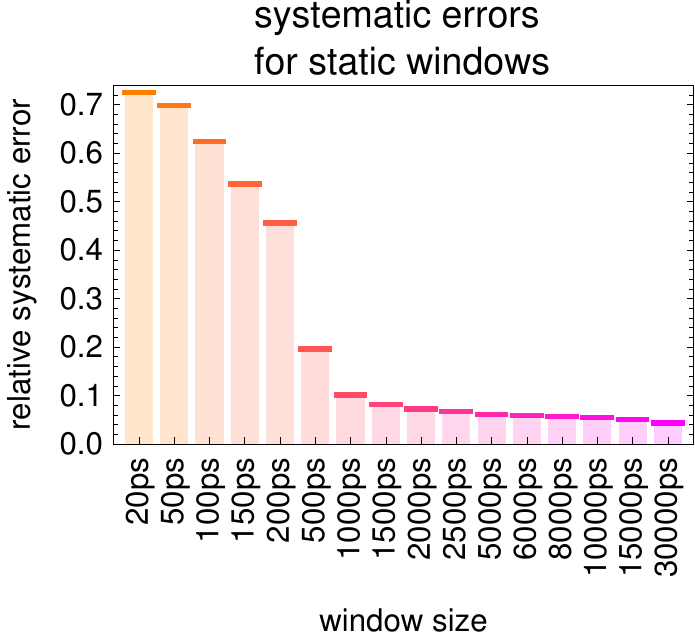}
	\hfill
	\includegraphics[width=.475\columnwidth]{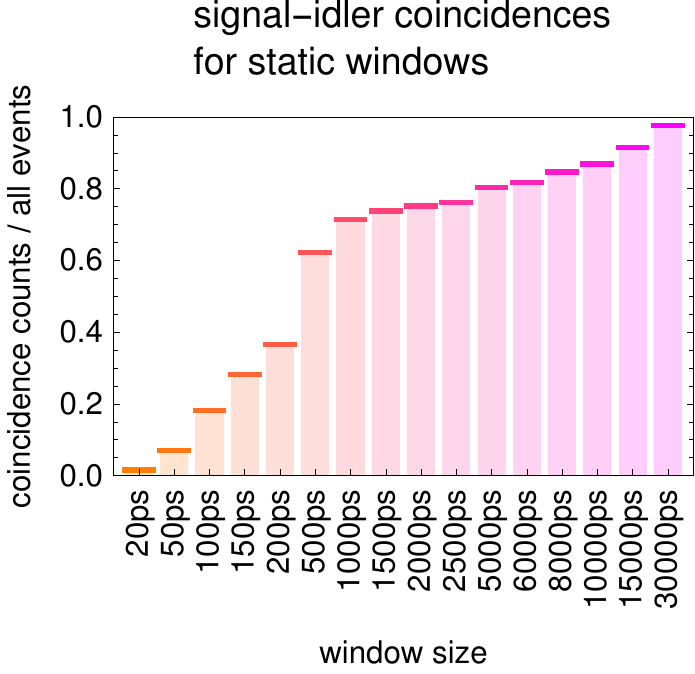}
	\\\vspace{2ex}
	\includegraphics[width=.475\columnwidth]{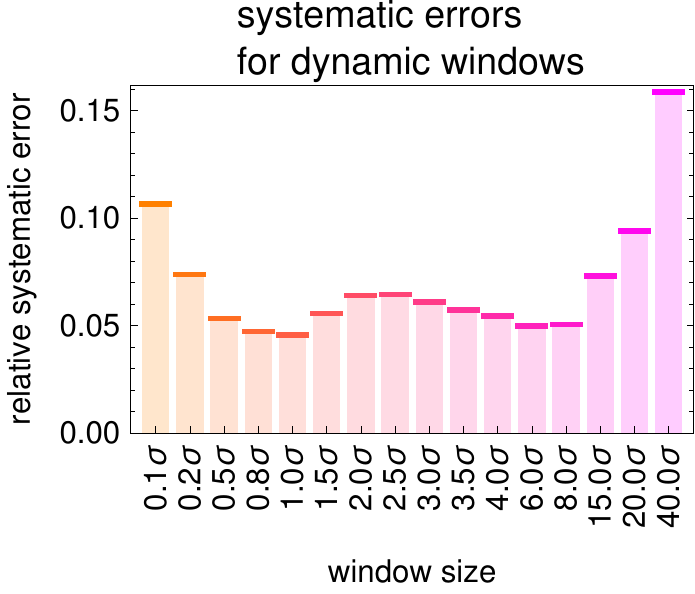}
	\hfill
	\includegraphics[width=.475\columnwidth]{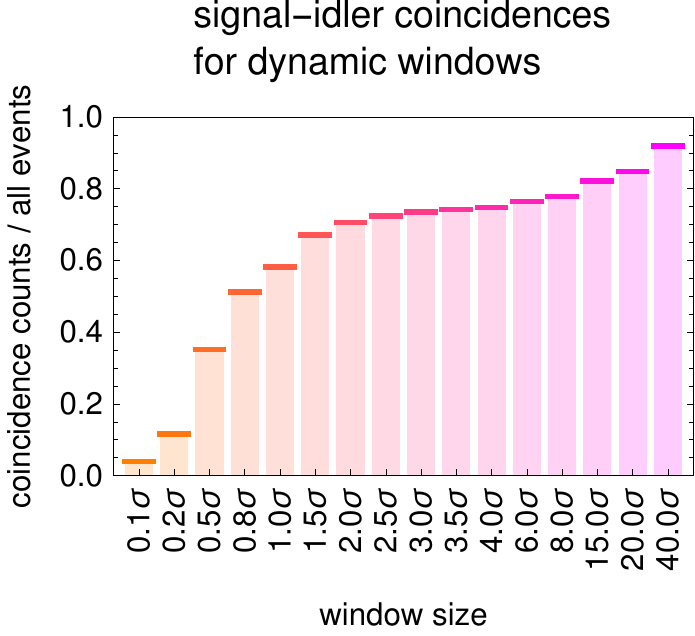}
	\caption{
		Systematic errors (left column) and coincidence counts (right column) for static windows (top row) and dynamic windows (bottom row) for a pump power $P=500\,\mathrm{\mu W}$.
	}\label{supp.fig:StaticVsDynamic}
\end{figure}

	In the left column in Fig. \ref{supp.fig:StaticVsDynamic}, systematic errors from asymmetries are analyzed for different binnings for static and dynamic windows in the top and bottom plot, respectively.
	Whilst the systematic error is roughly constant for the dynamic binning in a sensible range, e.g., $0.5\sigma$--$8\sigma$, the static case exhibits an increasingly larger systematic error when the window width is decreased.
	This effect can be explained by the fact that fluctuations over the detection bins (see Sec. \ref{supp.sec:SystematicError}) are much more pronounced when the window is narrower and an increasingly larger contribution of uncorrelated background noise for wider windows leads to diminished fluctuations, yet on the expense of the significance of certified nonclassicality (cf. Sec. \ref{supp.sec:JointCorrelations}).

	In the right column of Fig. \ref{supp.fig:StaticVsDynamic}, the ratio of coincidences and all counts, $\sum_{n_A,n_B\geq1}C(n_A,n_B)/C$, is shown for static (top) and dynamic (bottom) bins.
	In the limit of large windows, both exhibit the same asymptotic behavior since the background becomes the dominant contribution in both cases.
	For narrow windowing, however, coincidences are significantly pronounced for the dynamic case because it considers each bin on the basis of the underlying dispersion (i.e., pulse width broadening) on the same footing, which is not the case when it comes to correlating short and long optical wave packets in a static coincidence window.\\

\section{Nonclassicality of heralded states for different pump powers and dynamic window widths}\label{supp.sec:HeraldedSignificances}

\begin{figure*}
	\includegraphics[width=.99\textwidth]{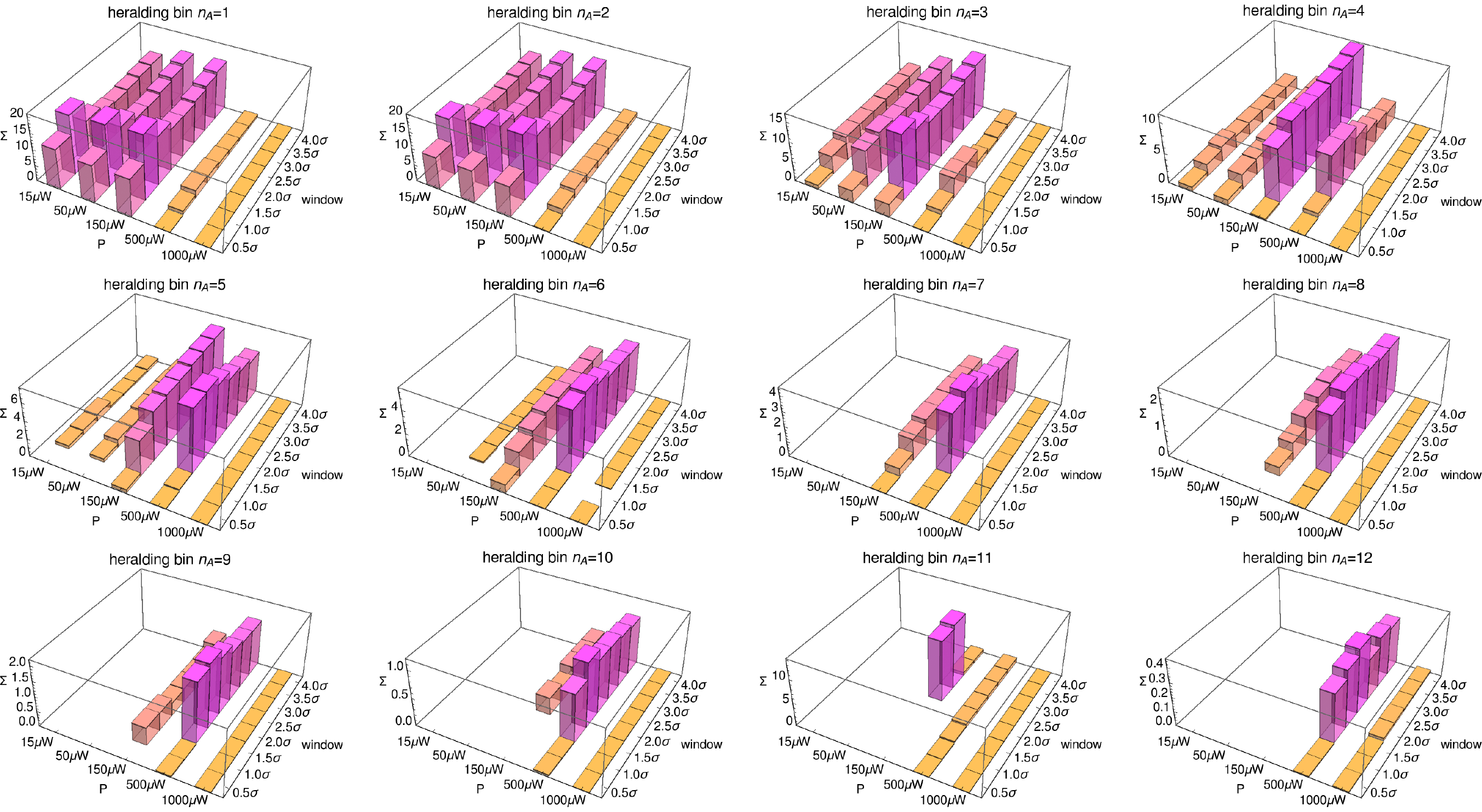}
	\caption{
		Significance $\Sigma$ of verified nonclassicality for heralding of $n_A$ photons in $A$ as a function of the pump power $P$ and size of dynamic coincidence windows.
		Nonclassicality is not detectable for parameter pairs for which no significance is depicted.
	}\label{supp.fig:StaticVsDynamicA}
\end{figure*}

\begin{figure*}
	\includegraphics[width=.99\textwidth]{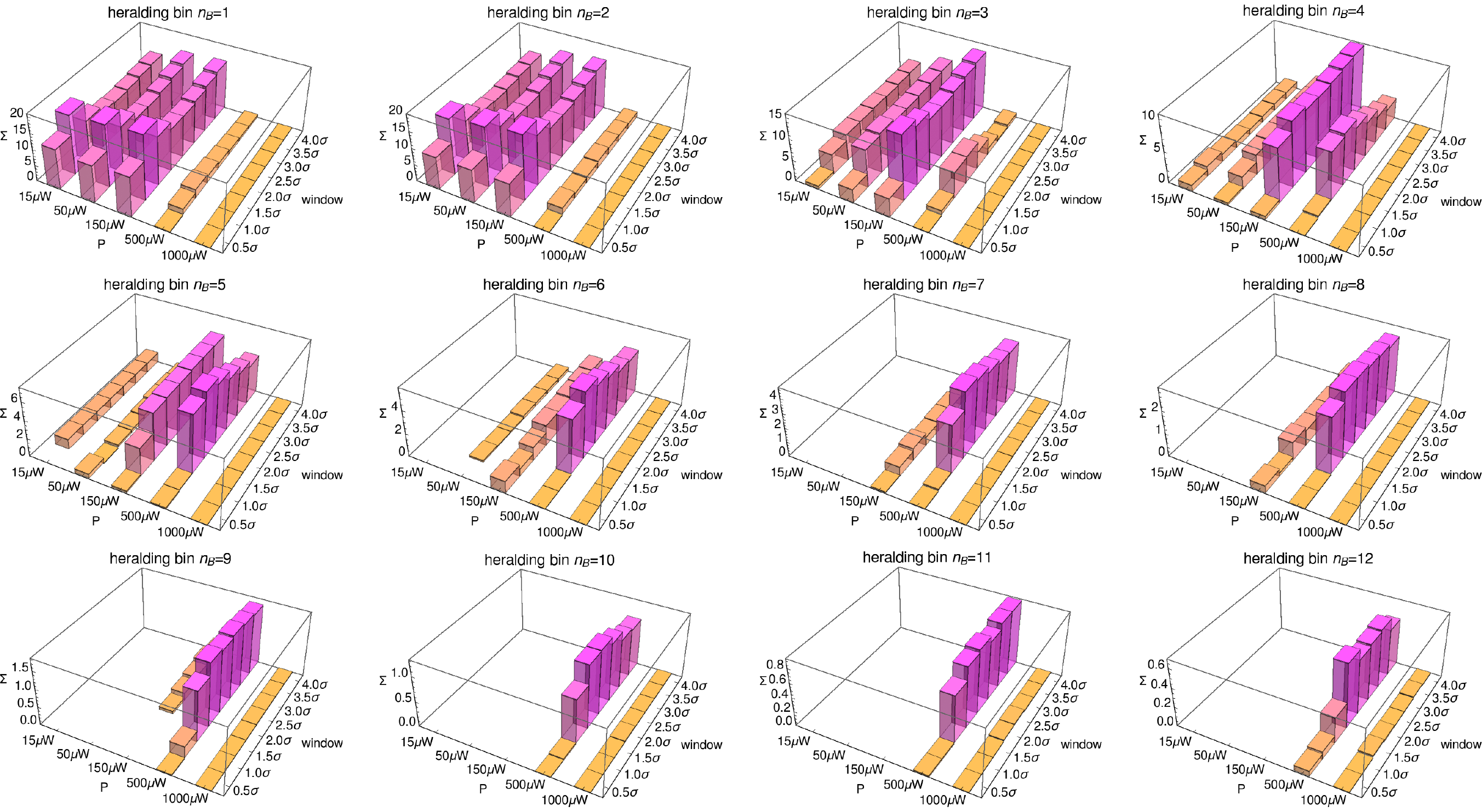}
	\caption{
		Significance $\Sigma$ of verified nonclassicality for heralding of $n_B$ photons in $B$ as a function of the pump power $P$ and size of dynamic coincidence windows.
		Nonclassicality is not detectable for parameter pairs for which no significance is depicted.
	}\label{supp.fig:StaticVsDynamicB}
\end{figure*}

	Figures \ref{supp.fig:StaticVsDynamicA} and \ref{supp.fig:StaticVsDynamicB} depict the nonclassicality analysis of heralded states---i.e., the single-mode nonclassicality when conditioned to the number $n$ of clicks in $A$ and $B$, respectively---for all measured pump powers, for various dynamic windows, and for up to twelve clicks from the herald detection.
	That is, the conditional click counts $C(n_B|n_A)=C(n_A,n_B)/C(n_A)$ and $C(n_A|n_B)=C(n_A,n_B)/C(n_B)$, where $C(n_A)=\sum_{n_B}C(n_A,n_B)$ and $C(n_B)=\sum_{n_A}C(n_A,n_B)$ are marginals, are used to construct a $65\times 65$ matrix of moments (i.e., up to $128$th order of single-mode moments).
	The resulting negativity in terms of the minimal negative eigenvalue of this matrix is normalized to the error estimate (including systematic and random errors) of that eigenvalue.
	This results in the shown significance $\Sigma$ of verified nonclassicality.
	A larger $\Sigma$ means a statistically higher certification of nonclassicality; no nonclassicality could be detected for blank entries in the plot.
	It is noteworthy that $960$ different scenarios are analyzed in Figs. \ref{supp.fig:StaticVsDynamicA} and \ref{supp.fig:StaticVsDynamicB}.

	As one might expect, for low photon numbers, the most robust nonclassicality is achieved for lower photon numbers because the heralding process leads to smaller higher photon-number contributions which could spoil the photon's purity.
	Conversely, higher pump powers are required to collect enough data to carry out the heralding to higher photon numbers in a statistically meaningful manner.
	Note that the symmetry of $A$ and $B$ is underlined by the similarity of the results when exchanging $n_A$ in Fig. \ref{supp.fig:StaticVsDynamicA} with $n_B$ in Fig. \ref{supp.fig:StaticVsDynamicB}.


